# More (or Less) Economic Limits of the Blockchain[1]

Joshua S. Gans (University of Toronto & NBER)
and Neil Gandal (Tel Aviv University & CEPR)

27th November, 2019

This paper extends the blockchain sustainability framework of Budish (2018) to consider proof of stake (in addition to proof of work) consensus mechanisms and permissioned (where the number of nodes are fixed) networks. It is demonstrated that an economically sustainable network will involve the same cost regardless of whether it is proof of work or proof of stake although in the later the cost will take the form of illiquid financial resources. In addition, it is shown that regulating the number of nodes (as in a permissioned network) does not lead to additional cost savings that cannot otherwise be achieved via a setting of block rewards in a permissionless (i.e., free entry) network. This suggests that permissioned networks will not be able to economize on costs relative to permissionless networks.

---

[1] We are grateful to Orbs for an academic grant and to Eric Budish and Oded Noam for helpful discussion, suggestions and comments that significantly improved the paper. All opinions (to the extent that they are opinions) are those of the authors.

# 1    Introduction

Blockchains rely on the validation of blocks (or transactions) using 'consensus mechanisms.' These mechanisms involve distributed nodes 'agreeing' as to whether a block is valid and whether it should be added to a chain. However, in order to prevent attacks, some cost must be placed on becoming a node in the network; critically, there must be a cost in proposing a block to add to the chain.

At present, the main consensus mechanisms are based on proof of work (PoW). This involves a cost of being a proposer in terms of real resources. In the Bitcoin protocol, for example, to be a proposer requires winning a computational game. The prize for winning is a block reward and transaction fee. The former is set by protocol and, if it is in cryptocurrency, the value of the currency. The latter is often set by users of the network. The cost of the contest is performing the computational task – that is, having computer hardware and energy resources.

PoW schemes have been criticized for the high level of energy resources required. In the Bitcoin network, these have been estimated to exceed the energy requirements of small countries. For this reason, alternative ways of imposing costs on becoming proposers have been explored.

The main alternative is proof of stake (PoS). This involves proposers proving their 'worthiness' by holding a stake – in terms of cryptocurrency – in the network. It is costly because they essentially make their holdings illiquid.[2]

The economic question is whether PoS type systems can perform more efficiently than PoW systems. We show, using the methodology for examining blockchain sustainability developed by Budish (2018), that the (perhaps) surprising answer is *no*! In the case of Permissionless blockchains (i.e. free entry,) the cost of PoW schemes are *identical* to the cost of PoS schemes.

We then examine Permissioned blockchains to see whether they can result in lower resource costs than Permissionless blockchains. The answer depends on whether the block reward is an exogenous or endogenous variable. If the block award is exogenous, then under certain conditions, Permissioned blockchains have lower costs than Permissionless blockchains. However, importantly, when the block reward is endogenous, the cost of Permissioned blockchains is again *identical* to that of Permissionless blockchains. We discuss the implications of this result in the conclusion.

---

[2] Proponents of PoS argue that security is achieved not only through incentive alignment, but also because attackers need to be significant stakeholders and as such are required to have "skin in the game." Thus, the argument goes, they would be reluctant to attack the system. In practice, this consideration may be important, but in the model we develop, if attackers gain more from the attack than they lose from the devaluation of their stake, they will indeed attack.

## 2      The Economics of Permissionless Blockchains

In a Permissionless blockchain, 'hostile nodes' have to harness/acquire enough computing power in order to 'attack' and fork the blockchain. The PoW protocol that was developed for Bitcoin has proven robust to various attacks but has shown difficulties in allowing potential upgrades in the operation of its blockchain.

While Permissionless blockchains facilitate exchange on decentralized systems without central authorities, they are not "tamper proof boxes." There are postulated means by which they could be disrupted – most notably through what is called a 'majority attack' whereby one entity controls a majority of nodes and can fork the blockchain according to their own preferences. Budish (2018) shows that fears of disruption have a strong basis. That paper examines when Bitcoin and other cryptocurrencies (i.e., digital currencies) using PoW would be vulnerable to being hijacked. Budish develops an equilibrium model that includes the (i) mining game (i.e., the supply side,) and (ii) incentive compatibility (the demand side). He concludes that Bitcoin "would be majority attacked if it became sufficiently economically important." (Budish, 2018, abstract)

In related work, Biais et.al. (2018) examine the "blockchain folk theorem." Their paper demonstrates that miners have an incentive to coordinate on the blockchain they are working on. At the same time, if a sufficient number of miners chose to do so, the same incentive to coordinate can lead them to shift quickly to a fork and work on that. In other words, coordination does not necessarily imply stability. Forks are likely to be driven by information delays and also software upgrades.[3] Similarly, Barrera and Hurder (2018) consider coordination as a driving force in blockchain stability. For a hard fork to arise, a large number of miners need to back the fork even in situations where not all miners follow. The basic model is similar to 'one CPU, one vote.' However, they argue that governance rules can smooth the process and lead to compromise solutions that would maximise the surplus of the community.[4]

---

[3] In economics, a folk theorem is a theorem many believe to hold even if a perfect proof of that theorem does not exist. In this case, it is the belief that miners will have incentives to work on one blockchain and that this will be the blockchain with the largest number of verified blocks.

# 3 A Simple Model of Sustainability using PoW: Permissionless Blockchain

In this paper, we expand the work of Budish (2018) to consider PoS systems and permissioned blockchains. But first we review his model for permissionless PoW blockchains. In doing this, we set up the conditions of sustainability so as to extend them to PoS as well as permissioned networks.

The PoW protocol requires that nodes perform a computational puzzle in order to add a valid block to the chain of transactions. To simplify matters, we suppose the following throughout the analysis:

- Blockchain nodes have a dollar cost, $c$, of competing for a block reward (or transaction fee). In the case of PoW, this is essentially the marginal cost of electricity plus equipment rental.
- The block reward is paid out as $P$ tokens.
- The exchange rate of tokens to dollars is $e$. Thus, the dollar block reward is $eP$.

## 3.1 Free Entry

A permissionless network implies that anybody can become a node (or miner) and will do so if it is profitable. If there are $N$ nodes with identical costs, $c$, the probability that a node earns the block reward is $1/N$. This means that a node's expected payoff is $\frac{1}{N}eP - c$.

Assuming that all nodes are honest, their number ($N$) in equilibrium is determined by a free entry constraint whereby the equilibrium number of nodes, $\hat{N}$ will be the highest $N$ such that $\frac{1}{N}eP \geq c$. Ignoring integer constraints, this implies:

$$(\text{FE}) \quad c = \frac{eP}{\hat{N}}$$

---

We will denote this condition as "free entry" (FE).

The (FE) condition has a strong implication for the resource cost per block, $Nc$: it is no higher than $eP$. This means that a fall in $c$ (possibly caused by a reduction in electricity prices or a choice to reduce the difficulty of computational puzzles) will not reduce the economic cost of mining since $N$ will rise. That is, given $P$ and $e$, $Nc$ is constant. This constraint arises in more detailed models where the mining process is modelled as a contest and where the difficulty of the computational task requirements adjusts as more computing power comes into the market (e.g., Ma, Gans and Tourky, 2019).

*3.2    Incentive Compatibility*

The (FE) condition dictates what drives miners to enter when they are 'honest' in the sense of being interested in processing transactions and validating blocks. However, miners could also be 'dishonest' in the sense of having other goals that cause them to want to append blocks with information they know to be false (e.g., as might arise in a double spend attack or in an attempt to sabotage the network for other reasons). A sustainable blockchain has to be robust against such agents; deterring their entry. That is, we have to check whether the PoW protocol implies that dishonest miners will have no incentive to enter. Budish (2018), the first researcher to consider this, terms this the *incentive compatibility* or (IC) constraint.

To derive this constraint, assume that there are $N$ honest miners. Conducting activities that are dishonest requires effective control of the network. At a minimum, in PoW, this requires a dishonest miner to control a majority of computing capacity – specifically, they need to add computing power equivalent to $N + \varepsilon$. This means that the cost of conducting dishonest activities on the network is at least $Nc$ per block.

One example of this is forking a network beyond a certain point. This involves mining a private network for a certain period of time until it has the longest chain (which is the coordinating device most used to form consensus on most PoW networks) and then making the network public attracting other miners. It is the fact of more computing power being applied to the private network by a dishonest miner that means that it will eventually generate the longest chain and be able to 'infect' the primary blockchain with its fork. This procedure is a precursor to a dishonest miner engaging in multiple spends on cryptocurrency.

Budish (2018) considers two limiting factors on a simple majority attack. First, some activities from dishonest miners may require more than a simple majority to implement. For instance, control to achieve a fork may require control of $\frac{A}{A+1}$ percent of the nodes. Thus, the cost per block for entry by a dishonest miner would be $ANc$ (where $A >$ 1).

Second, for some activities that involve interaction outside the blockchain (such as a multi-spend attack), control of the blockchain cannot be confined to just the block in question but may require a time period to elapse. Thus, the dishonest node may have to control the network for a time which translates into adding $t$ blocks.

Offsetting these limiting factors is the fact that, while controlling a network, a block reward ($eP$) that will be earned for each block added. That reward accrues to the dishonest miner. Putting these together, the net cost to the dishonest miner is $(ANc - eP)t$. The entry decision of a dishonest miner will be driven by the benefits they receive from such control – that is, from dishonest activities. Suppose that the private benefit for an attacker is $V(e)$.

Given these costs and benefits to a dishonest miner, we can see that dishonest entry to a blockchain network consisting of $N$ nodes will not be profitable if:

(IC) $\quad AtNc - teP \geq V(e) \Longrightarrow c \geq \frac{V(e) + teP}{AtN}$

We call this the (IC) condition and it comes from Budish (2018.) The left-hand side ($AtNc - teP$) is the cost of controlling $NA$ nodes for $t$ periods less the block reward earned during the control period. The right-hand side, $V(e)$, is the benefit of exercising that control for personal benefit. In the analysis that follows, we suppose that the private benefit for an attacker is $V(e)$; a non-decreasing function; that is, the more valuable is cryptocurrency, the greater is the private benefit from dishonest activities.

### 3.3    Stability

To summarize, the (FE) condition describes how many nodes will be established by miners for a given cost, $c$, involved in PoW while the (IC) constraint describes how high that cost, $c$, has to be to deter entry by dishonest miners. The challenge for sustainability arises because there is instrument in the design of the PoW protocol that can be used to guarantee that both the (FE) and (IC) conditions to be simultaneously satisfied.

To explore this, note that there are two parameters that can be embedded into a blockchain protocol ($c$ and $P$) that can impact on both the FE and IC conditions. Let's begin with $c$ – which can be set by changing the nature and likely energy requirements of any computational puzzle. Figure One shows what happens if $c$ is chosen to be a specific level, such as $c^*$. Note that the FE condition implies that, given this choice, the number of (honest) nodes will be $N^*$. As plotted, at this $N^*$, the IC constraint is satisfied. In other words, the blockchain is sustainable and, moreover, it is sustainable for a large range of $c$ because the FE constraint lies above the IC constraint.

Figure One

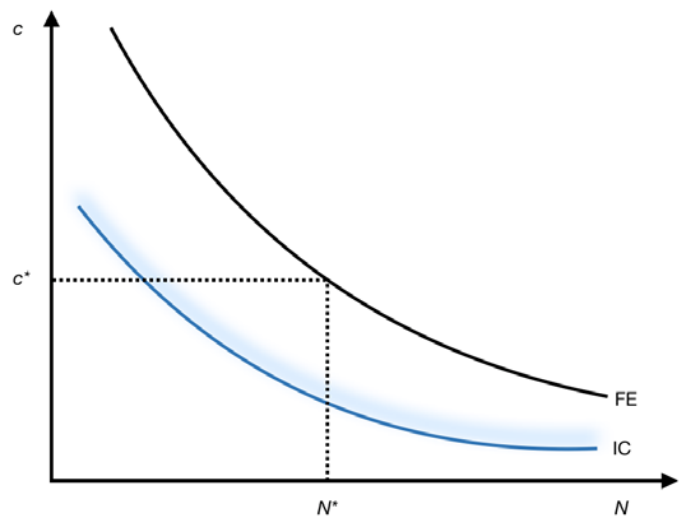

By contrast, in Figure Two, the IC constraint is above the FE condition. In this case, for any choice of $c$, the number of honest nodes that would enter (as implied by the FE condition) would be too few and would create an incentive for dishonest miners to enter. Thus, in this case, there is no value of $c$ that would make the blockchain sustainable.

Figure Two

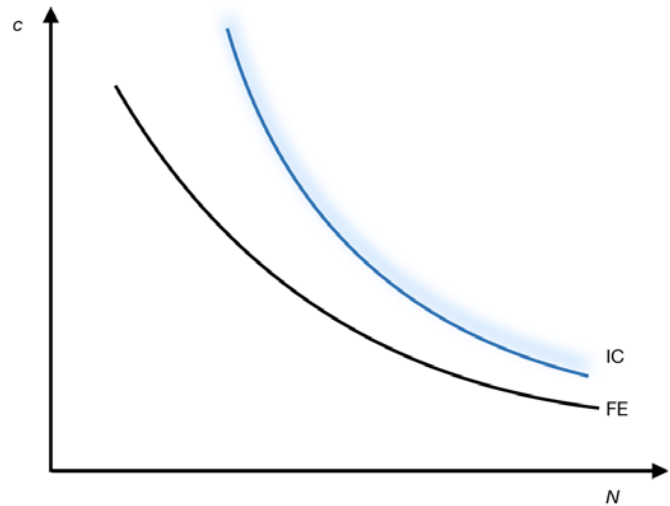

The reason for this can be seen by noting that both the FE and IC conditions are, in effect, defined relative to the total cost of PoW. The FE condition implies that $Nc = eP$ while the IC constraint requires that $Nc > \frac{V(e) + ePt}{At}$. Notice that the right-hand sides of these conditions do not depend on $N$ or $c$. Thus, sustainability is determined solely by whether:

$$(B) \qquad (A - 1)ePt > V(e)$$

This is the stability condition derived by Budish (2018); we refer to it as the Budish (B) condition. If $(A - 1)ePt < V(e)$, there is no permissionless system that satisfies the IC constraint. In other words, the only way mining pays is if there is a private benefit. Thus, we can conclude the following:

- If (B) holds, the FE condition lies above the IC constraint and the permissionless blockchain is stable for all values of $c$.

- If (B) does not hold, the FE condition lies below the IC constraint and the permissionless blockchain is not stable for any values of $c$.

Thus, the ability to control $c$ does not matter for stability. This is determined by other factors.

If $c$ cannot be used to create a sustainable permissionless blockchain, what about $P$? Note that, as $P$ increases, both the IC and FE conditions shift upwards. There are more nodes in equilibrium on the network and a higher aggregate cost ($Nc$). However, when $P$ rises, the net effect is to increase the FE curve by more than IC curve and promote sustainability.[5] Thus, we can conclude that, other things being equal, as $P$ rises, the likelihood of stability is increased.[6]

In designing a permissionless blockchain (and assuming $c$ is exogenous), the minimum block reward that will create a sustainable outcome is $P = \frac{V(e)}{e(A-1)t}$. Thus, so long as the block reward can be adjusted to 'market' conditions (notably $e$), it is possible to create a sustainable blockchain.

This result says that the block reward should change as the exchange rate, $e$, changes. But in which direction should the change go? Note that, as $e$ increases, the impact on sustainability depends on the relationship between $e$ and $V$. If $V$ is non-decreasing in $e$, the effect of an increase in $e$ depends upon the sign of $(eV'(e) - V(e))/e^2$ or whether the elasticity of $V$ with respect to $e$ is greater than 1. In this case, a higher value for the cryptocurrency, increases the probability that the blockchain will be vulnerable to attack. Therefore, if $e$ rises, to ensure that the blockchain is sustainable, $P$ has to rise as well if the elasticity of $V$ with respect to $e$ (that is, $\frac{e}{V(e)}V'(e)$) is greater than unity implying that an increase in $e$ causes a higher proportionate increase in $V$.

To put this another way, for Bitcoin, $P$ is set to fall over time as part of the protocol while $e$ is expected to be higher (at least by bitcoin enthusiasts).[7] The only way these two changes would imply long-term sustainability of the blockchain is if $\frac{e}{V(e)}V'(e) < 1$ or that the private payoff from an attack rises proportionately less than an increase in the value of bitcoin. Note that for multiple-spend attacks, this may be a reasonable assumption as the scale of such attacks is capped by real world motivations.[8] For other issues (such as

---

[5] This is because the partial derivative of $N$ with respect to $P$ from the FE equation is e/c, while the partial derivative of $N$ with respect to $P$ from the incentive compatibility constraint (when it binds with equality) is $e/(Ac,)$ where $A > 1$.

[6] Thus the prize "$P$" is not just an award for winning the tournament. It is also the "security cost" required to insure stability of the ecosystem.

[7] We have not explicitly addressed transaction fees in the analysis, but it is straightforward to include them. Thus "$P$" can be thought of as a combination of the block award and the transaction fees. The analysis is exactly the same, since dishonest miners will also receive the transaction fees when they successfully fork the blockchain. Over time, the block awards will decline. In order to insure that (B) holds, the transaction fees will have to be raised over time. Thus, as Budish points out, higher transaction fees will be necessary over time as the block reward falls.

[8] For instance, for a double-spend attack, the value of a transaction will drive incentives to attack the network and, in effect, defraud a 'real world' payor. For instance, at a Bitcoin exchange rate of $10,000 ($= e$), a simple majority value for $A$ ($= 1.01$) and a block reward of 12.5 BTC, then, if the required time is 6 hours (so assuming 10 minutes per block, $t = 36$), the value ($V$) must be greater than $45,000 to be worthwhile. If the

sabotage of the network), that may not be the case and Bitcoin has to be robust to them all. Without more knowledge regarding $V$, it is not possible to assess whether these changes in the block reward make Bitcoin more or less sustainable.

## 4 Proof of Stake

PoS protocols are an attempt to allow for consensus mechanisms without relying on real resources (as in PoW).[9] This is achieved by requiring nodes to *stake* a sufficient quantity of tokens in order to be considered as a validator for a new block of transactions. There are, however, different ways in which validator nodes are selected.

One class of methods is *chain-based*. In that method, a validator is chosen at random from nodes that hold the requisite stake. In effect, this means that validators have a probability of proposing a block (and receiving a block reward) based on the amount they have staked to the network. Like PoW, it typically takes some time (in terms of $t$ blocks) before a block is treated as final and relied upon.[10]

In a PoS network, suppose that $S$ tokens are required for a stake and that the dollar interest rate is $r$, then (in terms of our PoW notation) $c = reS$; that is, the opportunity cost per period of resources a node must hold to be 'staked' (e.g., the lost earnings on fiat currency that is held in tokens). The stake, $S$, can be chosen in a permissionless PoS protocol which means that, potentially like $c$, it is a design decision.

This allows us to write a (FE) condition for a PoS network. Like in a PoW network, any node has a probability $1/N$ of earning a block reward. Thus, the expected per block payoff to a node is $\frac{1}{N}eP - reS$. Thus, the (FE) condition becomes (ignoring integer constraints):

$$\text{(FE)'} \quad S = \frac{P}{rN}$$

---

[9] required time is just 1 hour ($t = 6$,) that threshold falls to \$7,500. It is this type of calculation that caused Budish (2018) to conclude that Bitcoin may not be sustainable.

[9] Formally, PoW and PoS are "Sybil" control mechanisms, rather than consensus protocols. These mechanisms need to be combined with consensus protocols to make the system work. For example, in the case of Bitcoin, the longest chain in the blockchain is the consensus rule.

[10] Another class of methods is based on Byzantine Fault Tolerance. In these methods, a node is chosen at random to be a validator but a block is only considered final if a supermajority ($A$) other staked nodes agree that it is valid. The advantage is that the block can be relied upon without having to wait $t$ periods of time. In order to compare PoS and PoW, we do not examine this alternative in the paper.

Note that, unlike PoW, the (FE) condition does not depend on the exchange value ($e$) of tokens. However, like PoW, the total network 'cost' (that is, $SrN$) is fixed – that is, an increase in $S$, causes a proportionate reduction in the number of nodes ($N$).

What about protection against attacks by dishonest nodes? Both PoS methods are vulnerable to attack forms based on establishing a private chain with altered transactions before releasing to publicly. With PoW, this entails a cost as a dishonest node is required to perform the PoW of the entire network in order to obtain the longest chain upon publication. With PoS, there is no such cost. The main challenge comes, however, that when the alternative chain is published there is the challenge of getting other nodes to accept it. For nodes that were online while the alternative chain was being written, they will be able to identify the alternative chain. For new nodes or ones that were offline, they cannot tell which is the legitimate chain. Thus, for an attack to be successful, the dishonest node needs to take actions that would shift the share of online versus other nodes. We assume that this takes time ($t$ periods.)

Such attacks rely on the attacker building on both the main chain and their alternative at the same time. This is something that is possible with PoS but costly for PoW. However, networks have implemented various methods to guard against this. One such method is called 'slashing.' This involves the stake of a node being reduced or destroyed if it is found that they have worked on multiple chains. This is something that can be algorithmically detected.

That said, while such methods can prevent 'low scale' attacks on the network, PoS networks are still vulnerable to a majority-attack – as we examined for PoW. Such an attack requires the attacker to stake a supermajority of nodes for $t$ periods. If the value of an attack is $V(e)$ as before, then the cost of the attack is $ANtreS$ less the block reward $teP$ earned on the alternative chain. Note that block reward accrues to the attacker precisely because slashing or other mechanisms penalizes others if they work on the alternative chain leaving all of the block rewards to the attacker.

Given this, the (IC) constraint is

$$\text{(IC)'} \quad ANteSr - teP \geq V(e) \implies S \geq \frac{V(e) + teP}{eANtr}$$

Thus, so long as the stake, $S$, is sufficiently high, an attack can be prevented.

We can now perform a comparable exercise to that for PoW to examine what will determine the sustainability of a PoS blockchain. In particular, using (FE)' and (IC)', if the (IC)' constraint is to be satisfied while the equilibrium number of nodes is determined by the (FE)' condition that requires:

$$(B)' \quad (A-1)teP \geq V(e)$$

Note, critically, that this condition is *identical* to the PoW Budish condition (B). Moreover, it is independent of the level of the stake (*S*).

In other words, despite the ability to control *S*, **there are no levers under proof of stake that will lead to greater sustainability than under proof of work**. In fact, in designing a permissionless blockchain (even though *S* can be chosen), the minimum block reward that will create a sustainable outcome is $P = \frac{V(e)}{e(A-1)t}$. This also means that the same elasticity condition on *V*(*e*) drives whether, for a fixed block reward, the network will be more sustainable as *e* grows.

It is useful to note, however, that the mechanism is different. In particular, under PoS, the FE condition is independent of the exchange rate (*e*). In other words, the number of nodes will not change as the exchange rate changes and will be pinned down by the level of the stake. Thus, the size of the permissionless network can be controlled by changing the stake. This is not the case in PoW, since given *P* and *e*, *c* (which is a design variable) determines the network size (*N*) in the permissionless PoW mechanism.[11]

## 5    Permissioned Blockchains

PoW and PoS involve the same 'dollar' cost in running under permissionless protocols. If the goal was to set up a sustainable blockchain at a lower cost, this suggests that a permissioned network that regulated the number of nodes (*N*) might be a means of lowering *cN*; the total cost of running the network. Here we consider what happens when *N* (in addition to *c*, *S*, and *P*) can be chosen. We will not presume anything about the selection process for node operators – that is, none of them will presumed to be trusted – and, thus, participation and incentive constraints will have to continue to hold.[12]

---

[11] Of course, it can be argued that in addition to the real resources used to sustain stability, PoW causes negative externalities. Using large amounts of electricity, most of which is not generated by renewable sources, but rather by methods (like burning coal) that cause pollution and perhaps global warming. Thus PoW mechanisms have higher total costs (internal plus external costs) than PoS mechanisms.

Given this, in a PoW blockchain, the protocol designer solves the following problem:

$$Min_{c,N,P} cN \text{ subject to } ANtc - teP \geq V(e) \text{ and } Nc \leq eP$$

The IC constraint, when it binds, pins down the aggregate cost at $\frac{V(e)+teP}{At}$. Note that this is increasing in the block reward ($P$). Thus, the designer will want to reduce the block reward. This reduction will be constrained by the FE condition (as it needs to be so that nodes are not making losses); that is, $P = \frac{1}{e} Nc = \frac{V(e)+teP}{eAt} \implies P = \frac{V(e)}{e(A-1)t}$ where the last substitution comes from the substitution of the IC constrained aggregate cost. This the same block reward as in a permissionless system and, thus, there or no aggregate cost savings from a permissioned system.

Intuitively, when the Budish condition (B) holds, the FE condition lies above the IC constraint and the question is whether, by choosing $N$, the network can operate on the IC constraint at a lower $cN$. This would give a total cost of $\frac{V(e)+teP}{At}$ which is increasing in $P$. Thus, the designer will want to reduce $P$ to be as small as possible. This is precisely what a designer would want to achieve if the network were permissionless and (B) held. If (B) does not hold, in both the permissioned and permissionless cases, the designer wants to increase $P$ so that the FE and IC constraints *both* bind.

Changing $N$ will not improve prospects for sustainability if (B) does not hold. In such a case, there is no ($c$, $N$) for which both the FE and IC constraints hold. It is only by increasing the block reward ($P$) and adjusting it to 'market' conditions that network costs can be minimized while allowing for a sustainable blockchain.

Under PoS, the network chooses ($S$, $N$) in addition to $P$. In this case, the IC constraint implies that total cost, $eSNr = \frac{V(e)+teP}{At}$. Once again, this shows that there are no gains from having a permissioned system.

---

[12] In practice, "permissioned blockchains" usually are private or consortium networks in which members are both limited in number and individually selected. Therefore the members' identities are known. This may generate additional security since "bad actors" may be liable under civil and criminal law. This effect is not included in the model, but it is an additional factor to consider as research on the topic moves forward.

Note that if $P$ is fixed, then regulating $N$ will reduce costs if (B) holds by allowing the network to operate on the IC rather than the *FE* constraint. The intuition can be seen by examining Figure One. In this case, the FE constraint lies above the IC constraint. (Recall that this does not depend on $(c, N)$). Since the FE lies above the IC constraint, we have sustainability for a Permissionless blockchain.

Now consider a Permissioned blockchain. For a given $c$, $N$ can be chosen so that the IC constraint binds exactly. This results in lower costs $cN$. Hence, a Permissioned blockchain can reduce the costs relative to a Permissionless blockchain, but only in the case in which $P$ is fixed.[13]

# 6      Conclusion

In summary, this analysis shows that there is no difference between proof of work and proof of stake in terms of resource cost. Further, when the block reward is endogenous, there is no difference in costs between Permissionless and Permissioned blockchains. Regardless of whether it is PoW or PoS or permissioned/permissionless, being able to have a block reward that can be adjusted to market conditions is the way to insure sustainable blockchains. The results of this analysis suggest that when the block reward is a choice variable, a hybrid system will not have lower resources cost than Permissionless on Permissioned blockchains.[14] Thus, while Permissioned and hybrid systems may be attractive for other reasons, cost savings is not one of them.

---

[13] Of course, $N$ cannot be chosen too small. Otherwise, the combination of $(c, N)$ leads us to a point below the IC curve. This bounds the cost saving from a permissioned blockchain.

[14] Several papers examine how such hybrid systems might work. See Pass and Shi (2017) and Jiang et al (2018) to see how such hybrid systems might work.